\documentclass[preprint,a4paper,superscriptaddress]{revtex4}
\usepackage{graphicx}

\begin{document}

\newcommand {\R}{{\mathcal R}}
\newcommand{\al}{\alpha}

\title{WMAP Constraints on varying $\alpha$ and the
Promise of Reionization}

\author{C.J.A.P. Martins}
\email[Electronic address: ]{C.J.A.P.Martins@damtp.cam.ac.uk}
\affiliation{Centro de Astrof\'{\i}sica da Universidade do Porto, R. das
Estrelas s/n, 4150-762 Porto, Portugal}
\affiliation{Department of Applied Mathematics and Theoretical Physics,
Centre for Mathematical Sciences,\\ University of Cambridge,
Wilberforce Road, Cambridge CB3 0WA, United Kingdom}
\affiliation{Institut d'Astrophysique de Paris, 98 bis Boulevard Arago,
75014 Paris, France}
\author{A. Melchiorri}
\email[Electronic address: ]{melch@astro.ox.ac.uk}
\affiliation{Department of Physics, Nuclear \& Astrophysics Laboratory,
University of Oxford,\\ Keble Road, Oxford OX1 3RH, United Kingdom}
\author{G. Rocha}
\email[Electronic address: ]{graca@mrao.cam.ac.uk}
\affiliation{Astrophysics Group, Cavendish Laboratory,
Madingley Road, Cambridge CB3 0HE, United Kingdom}
\affiliation{Centro de Astrof\'{\i}sica da Universidade do Porto, R. das
Estrelas s/n, 4150-762 Porto, Portugal}
\author{R. Trotta}
\email[Electronic address: ]{roberto.trotta@physics.unige.ch}
\affiliation{D\'epartement de Physique Th\'eorique, Universit\'e de
Gen\`eve, 24 quai Ernest Ansermet, CH-1211 Gen\`eve 4, Switzerland}
\author{P. P. Avelino}
\email[Electronic address: ]{ppavelin@fc.up.pt}
\affiliation{Centro de F\'{\i}sica do Porto e
Departamento de F\'{\i}sica da Faculdade de Ci\^encias
da Universidade do Porto,\\ Rua do Campo Alegre 687, 4169-007 Porto, Portugal}
\author{P.T.P. Viana}
\email[Electronic address: ]{viana@astro.up.pt}
\affiliation{Centro de Astrof\'{\i}sica da Universidade do Porto, R. das
Estrelas s/n, 4150-762 Porto, Portugal}
\affiliation{Departamento de Matem\'atica Aplicada da Faculdade de Ci\^encias
da Universidade do Porto,\\ Rua do Campo Alegre 687, 4169-007 Porto, Portugal}

\date{21 July 2003}

\begin{abstract}
We present up-to-date constraints on the value of the
fine-structure constant at the epoch of decoupling from the recent
observations made by the Wilkinson Microwave Anisotropy Probe
(WMAP) satellite. In the framework of models we considered, a
positive (negative) variation of the value of $\alpha$ at
decoupling with respect to the present-day value is now bounded to
be smaller than $2\% $ ($6\% $) at $95\% $ confidence level. We point out that
the existence of an early reionization epoch as suggested by the
above measurements will, when more accurate cosmic microwave
background polarization data is available, lead to considerably
tighter constraints. The so-called `reionization bump', in particular,
will be extremely useful for this purpose.
We find that the tightest possible constraint on $\alpha$ is
about $0.1 \%$ using CMB data alone---tighter constraints
will require further (non-CMB) priors.
\end{abstract}
\preprint{DAMTP-2003-13}
\maketitle

\section{Introduction}

Now that WMAP has validated beyond reasonable doubt the so-called
concordance (or is it conspiracy?)
model of cosmology \cite{Bennett,Hinshaw,Kogut,Verde},
it is time to proceed to the next level of questioning and start
asking what are the `dark components' that make up about $96\%$ of the
energy budget of the universe. Most of this is in some non-baryonic form,
for which there is at present no direct evidence or solid theoretical
explanation.
Understanding its nature will clearly require further progress in
fundamental physics.

Cosmology and astrophysics provide a laboratory with extreme
conditions in which to test fundamental physics and search for new
paradigms. Currently preferred unification
theories \cite{Polchinski,Thibault}
predict the existence of additional
space-time dimensions, which have a number of possibly observable
consequences, including modifications in the gravitational laws on
very large (or very small) scales \cite{Will} and space-time variations of the
fundamental constants of nature \cite{Uzan,Essay}. Recent evidence
of a time variation of fundamental constants
\cite{Webb,Jenam,Ivanchik} offers an important opportunity to test
such fundamental physics models. It should be noted that the issue
is not \textit{if} such theories predict such variations, but
\textit{at what level} they do so, and hence if there is
any hope of detecting them in the near future.

The most promising case is that of the fine-structure constant
$\alpha$, for which some evidence of time variation at redshifts
$z\sim2-3$ already exists \cite{Webb,Jenam}. Since one expects
$\alpha$ to be a non-decreasing function of time \cite{Damour,Santiago,Barrow}, it
is particularly important to try to constrain it at earlier
epochs, where any variations relative to the present-day value
should be larger. (On the other hand, local laboratory tests can
also provide useful constraints \cite{Marion}.) The cosmic microwave background (CMB)
anisotropies provide such a probe, being mostly sensitive to the
epoch of decoupling, $z \sim 1100$. Here we extend our previous
work in this area \cite{Old,Avelino,Martins} by analyzing the recently
released WMAP first-year data and providing updated constraints on
the value of $\alpha$ at decoupling. We emphasize that in previous work,
constraints on $\alpha$ were obtained with the help of additional cosmological
datasets. Here, by contrast, we will analyse the WMAP dataset alone.
We also discuss how these
constraints can be improved in the future, when more precise CMB
polarization data will be available. In particular, we show that
the existence of an early reionization epoch is of significant
help in further constraining $\alpha$. A more detailed analysis of
this issue can be found in a companion paper \cite{Rocha}.

\section{WMAP constraints on $\alpha$}

We compare the recent WMAP temperature and cross-polarization data
with a set of flat cosmological models adopting the likelihood
estimator method described in \cite{Verde}. The models are
computed through a modified version of the CMBFAST code with
parameters sampled as follows: physical density in cold dark
matter $0.05 < \Omega_ch^2 < 0.20$ (step $0.01$), physical density
in baryons $0.010 < \Omega_bh^2 < 0.028$ (step $0.001$), 
cosmological constant $0.0 < \Omega_\Lambda < 0.95$ step $0.05$,
$0.80 <\alpha_{\text{dec}} / \alpha_0 <1.18$ (step $0.02$). Here $h$ is
the Hubble parameter today, $H_0 \equiv 100h$ km s$^{-1}$
Mpc$^{-1}$ which is related to the above quantities by the flatness condition, 
while $\alpha_{\text{dec}}$ ($\alpha_0$) is the value
of the fine structure constant at decoupling (today). 
We verified that the marginalized likelihood distribution from WMAP 
for each of the above parameters has negligible values when
computed at the extrema of the range allowed in the database. 
We also vary the optical depth $\tau$ in the range 
$0-0.30$ (step $0.04$) and the scalar spectral index of primordial 
fluctuations $0.88 < n_s< 1.08$ (step $0.005$). 
The upper limit for the optical depth is well motivated
by numerical simulations (see e.g. \cite{ciardi}).

We don't consider gravity waves, running
of the spectral index or iso-curvature modes since these further
modifications are not required by the WMAP data.

\begin{figure}
\includegraphics[width=3.5in]{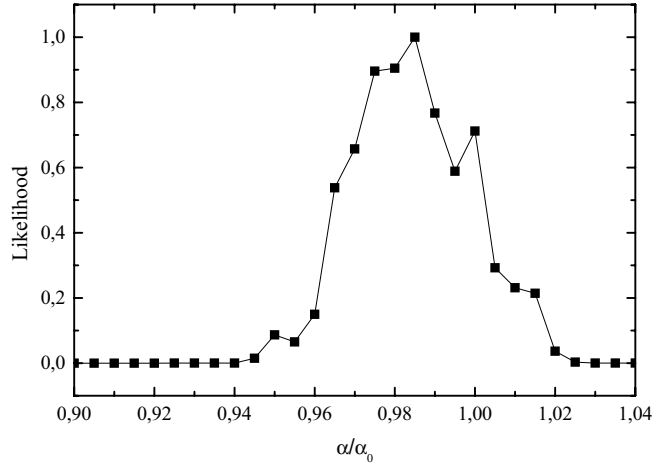}
\caption{\label{figalpha}
Likelihood distribution function for variations in the fine structure
constant obtained by an analysis of the WMAP data.}
\end{figure}

The likelihood
distribution function for $\alpha_{\text{dec}} / \alpha_0$,
obtained after marginalization over the remaining parameters, is
plotted in Figure \ref{figalpha}. We found, at $95 \%$ C.L. that $0.94 <
\alpha_{\text{dec}} / \alpha_0 < 1.01$, improving previous bounds
(see \cite{Martins} for the previous tighter bound, and
\cite{Old,Avelino,Battye} for earlier work) based on CMB and complementary datasets.
Thus  when it comes to constraining $\alpha$ in
the primordial universe, the WMAP dataset alone is at least as good
as everything else before it put together.

\section{The future: Polarization and reionization}

In our previous work \cite{Martins}, only the CMB temperature was
taken into account. The precision with which cosmological
parameters can be reconstructed using both CMB temperature and
E-polarization measurements is now re-examined by means of the
Fisher Matrix Analysis (FMA) technique. This technique was first applied
in this context by \cite{originalfma} (see also \cite{copyfma}).
We consider the planned
Planck satellite and an ideal experiment which would measure both
temperature and polarization to the cosmic variance limit (in the
following, 'CVL experiment`).

\begin{table}
\caption{\label{exppar} Experimental parameters for Planck
(nominal mission). After removal of the foregrounds, $f_{\rm sky}$ is
the effective fraction of the sky covered, while $\ell_{\rm max}$ is the largest multipole attained.} 
\begin{ruledtabular}
\begin{tabular}{|l|ccc|}
&  \multicolumn{3}{c|}{Planck HFI} \\
 \hline $\nu$ (GHz)  &
               $100$ &  $143$ & $217$  \\
$\theta_c$ (arcmin)&
    $10.7$ & $8.0$ & $5.5$ \\
$\sigma_{cT}$ ($\mu$K)  &
    $5.4$  & $6.0$  & $13.1$  \\
$\sigma_{cE}$ ($\mu$K)  &
    $n/a$  & $11.4$  & $26.7$  \\
$w^{-1}_c \cdot 10^{15}$ (K$^2$ ster) &
    $0.215$ & $0.158$ & $0.350$  \\
$\ell_c$ &
    $757$ & $1012$ & $1472$ \\\hline
$\ell_{\rm max}$ & \multicolumn{3}{c|}{$2000$} \\
$f_{\rm sky}$    & \multicolumn{3}{c|}{$0.80$} \\
\end{tabular}
\end{ruledtabular}
\end{table}

Cosmological models are characterized by the 8 dimensional
parameter set
\begin{equation}
{\bf \Theta} = (\Omega_b h^2, \Omega_m h^2, \Omega_\Lambda h^2,
\R, n_s, Q, \tau, \al),
\end{equation}
where  $\Omega_m = \Omega_c +\Omega_b$ is the energy density in
matter and $\Omega_\Lambda$ the energy density due to a
cosmological constant, and the Hubble parameter $h$ is a dependent
variable. The quantity $\R \equiv \ell_{\rm ref}  / \ell$ is the
`shift' parameter and \mbox{$Q = < \ell (\ell + 1) C_\ell
> ^{1/2}$} denotes the overall normalization, where the mean is
taken over the multipole range $2 \leq \ell \leq 2000$.
We assume  purely adiabatic initial conditions and we do not allow
for a tensor contribution. The Fisher matrix expansion is done
around a fiducial model with
parameters $\omega_b = 0.0200$, $\omega_m = 0.1310$,
$\omega_\Lambda = 0.2957$ (and $h = 0.65$), $\R = 0.9815$, $n_s =
1.00$,$Q = 1.00$, $\tau=0.20$ and $\al/\al_0=1.00$. In our
previous work \cite{Martins} the numerical accuracy of the FMA was
limited by the fact that differentiating around a flat model
requires computing open and closed models, which are calculated
using different numerical techniques. Here we instead
differentiate around a slightly closed model (as preferred by
WMAP) with $\Omega_{\rm{tot}} = 1.01$  to avoid extra sources of
numerical inaccuracies. We refer to \cite{Martins,Rocha} for a
detailed description of the numerical technique used. The
experimental parameters used for the Planck analysis are in Table
\ref{exppar}, and we use the first 3 channels of the Planck High
Frequency Instrument (HFI) only. Note that we account for possible
issues arising from point sources, foreground removal and galactic
plane contamination by assuming that once these have been taken into
account we are left with a `clean' fraction of the sky given by 
 $f_{\rm{sky}}$. For the cosmic variance limited
(CVL) experiment, we set the experimental noise to zero, and we
use a total sky coverage $f_{\rm{sky}} = 1.00$. Although this is
never to be achieved in practice, the CVL experiment illustrates
the precision which can be obtained {\it in principle} from CMB
temperature and E-polarization measurements.

\begin{figure}
\includegraphics[width=3.5in]{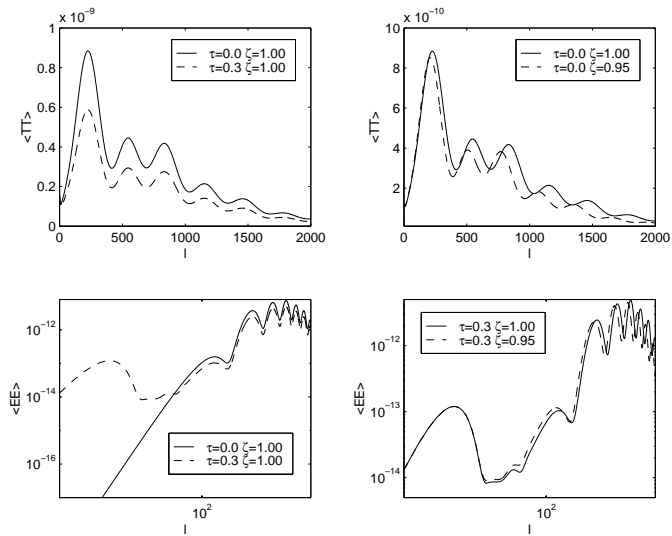}
\caption{\label{figcells} Contrasting the effects of varying
$\alpha$ and reionization on the CMB temperature and polarization.
Here $\zeta=\al_{dec}/\al_0$.}
\end{figure}

\begin{figure}
\includegraphics[width=3.5in]{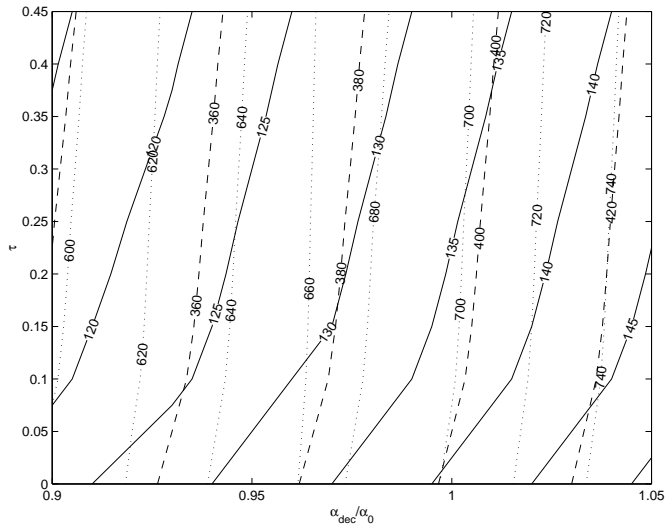}
\caption{\label{figpeaks} The separation in $\ell$ between the
reionization bump and the first (solid lines), second (dashed) and
third (dotted) peaks in the polarization spectrum, as a function
of $\alpha$ at decoupling and $\tau$. An instantaneous reionization
model is assumed.}
\end{figure}

Fig.~\ref{figcells} illustrates the effect of $\alpha$ and $\tau$
on the CMB temperature and polarization power spectra---see
\cite{Rocha} for a more detailed discussion. The CMB power
spectrum is, to a good approximation, insensitive to {\it how}
$\alpha$ varies from last scattering to today. Given the existing
observational constraints, one can therefore calculate the effect
of a varying $\alpha$ in both the temperature and polarization
power spectra by simply assuming two values for $\alpha$, one at
low redshift (effectively today's value, since any variation of
the magnitude of \cite{Webb} would have no noticeable effect) and
one around the epoch of decoupling, which may be different from
today's value. For the CMB temperature, reionization simply
changes the amplitude of the acoustic peaks, without affecting
their position and spacing (top left panel); a different value of
$\alpha$ at the last scattering, on the other hand, changes both
the amplitude and the position of the peaks (top right panel). The
outstanding effect of reionization is to introduce a bump in the
polarization spectrum at large angular scales (lower left panel).
This bump is produced well after decoupling at much lower
redshifts, when $\alpha$, if varying, is much closer to the
present day's value. We emphasize that, for the reason pointed out above,
\textit{we do not need specify a redshift dependence for the
variation}, although we could if we so choose. If the value of
$\alpha$ at low redshift is different from that at decoupling, the
peaks in the polarization power spectrum at small angular scales
will be shifted sideways, while the reionization bump on large
angular scales won't (lower right panel). It follows that by
measuring the separation between the normal peaks and the bump,
one can measure both $\alpha$ and $\tau$, as illustrated in Fig.
\ref{figpeaks}. Thus we expect that the existence of an early
reionization epoch will, when more accurate cosmic microwave
background polarization data is available, lead to considerably
tighter constraints on $\alpha$.

\begin{table}
\caption{\label{fmaresults}Fisher matrix analysis results for a
model with varying $\alpha$ and reionization: expected $1\sigma$
errors for the Planck satellite and for the CVL experiment (see
the text for details). The column {\it marg.} gives the error with
all other parameters being marginalized over; in the column {\it
fixed} the other parameters are held fixed at their fiducial value; in
the column {\it joint} all parameters are being estimated
jointly.}
\begin{ruledtabular}
\begin{tabular}{|c| c c c|c c c|}
 &  \multicolumn{6}{c}{$1\sigma$ errors (\%)} \\\hline
             & \multicolumn{3}{c}{Planck HFI} & \multicolumn{3}{c}{CVL} \\
             & marg.  & fixed & joint  & marg.  & fixed & joint           \\\hline
     & \multicolumn{6}{c}{E-Polarization Only (EE)} \\\hline
$\alpha$       &2.66       &0.06       &7.62     &0.40        &$<0.01$        &1.14 \\
$\tau$         &8.81       &2.78       &25.19    &2.26 &1.52 &6.45
\\ \hline
     & \multicolumn{6}{c}{Temperature Only (TT)} \\\hline
$\alpha$       &0.66       &0.02       &1.88     &0.41        &0.01        &1.18  \\
$\tau$         &26.93      &8.28       &77.02    &20.32 &5.89
&58.11 \\ \hline
     & \multicolumn{6}{c}{Temperature + Polarization (TT+EE)} \\\hline
$\alpha$       &0.34        &0.02      &0.97     &0.11        &$<0.01$        &0.32 \\
$\tau$         &4.48        &2.65      &12.80    &1.80       &1.48        &5.15 \\
\end{tabular}
\end{ruledtabular}
\end{table}

\begin{figure}
\includegraphics[width=3.5in]{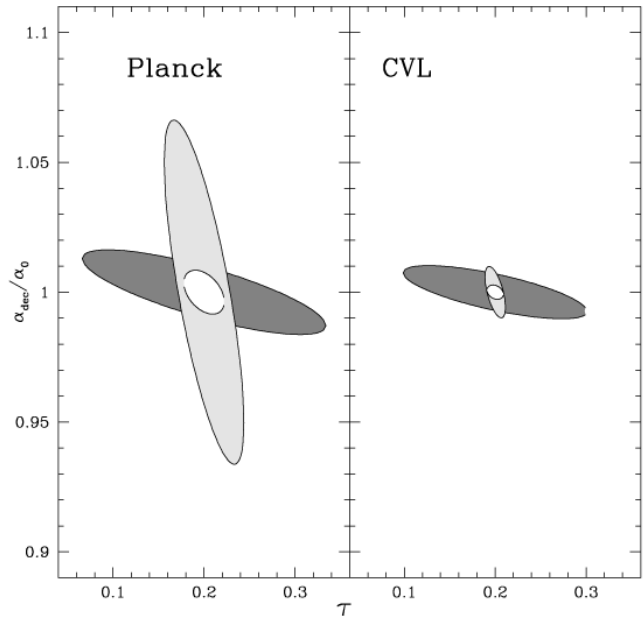}
\caption{\label{figlike} Ellipses containing $95.4\%$ ($2\sigma$)
of joint confidence in the $\alpha$ vs. $\tau$ plane (all other
parameters marginalized), for the Planck and cosmic variance
limited (CVL) experiments, using temperature alone (dark grey),
E-polarization alone (light grey), and both jointly (white).}
\end{figure}

Table \ref{fmaresults} and Fig.~\ref{figlike} summarize the
forecasts for the precision in determining $\tau$ and $\alpha$
(relative to the present day value) with Planck and the CVL
experiment. We consider the use of temperature information alone
(TT), E-polarization alone (EE) and both channels (EE+TT) jointly.
Note that one could use the temperature-polarization cross
correlation (ET) instead of the E-polarization, with the same
results. As it is apparent from Fig.~\ref{figlike}, TT and EE
suffer from degeneracies in different directions, because of the
reasons explained above. Thus combination of high-precision
temperature and polarization measurements can constrain in the
most effective ways both variations of $\alpha$ and $\tau$. Planck
will be essentially cosmic variance limited for temperature but
there will still be considerable room for improvement in
polarization (Table \ref{fmaresults}). This therefore argues for a
post-Planck polarization experiment, not least because
polarization is, in itself, better at determining cosmological
parameters than temperature. We conclude that Planck alone will be
able to constrain variations of $\alpha$ at the epoch of
decoupling within $0.34 \%$ ($1\sigma$, all other parameters
marginalized), which corresponds to approximately a factor 5
improvement on the current upper bound. On the other hand, the CMB
\textit{alone} can only constrain variations of $\alpha$ up to
${\cal O}(10^{-3})$ at $z \sim 1100$. Going beyond this limit
will require additional (non-CMB) priors on some of the other
cosmological parameters. This result is to be
contrasted with the variation measured in quasar absorption
systems by Ref.\cite{Webb}, $\delta \alpha / \alpha_0 = {\cal
O}(10^{-5})$ at $z \sim 2$. Nevertheless, there are models
where deviations from the present value could be detected
using the CMB.

\section{Conclusions}

We have tightened the CMB constraints on the value of the
fine-structure constant at the epoch of decoupling, using the WMAP satellite data.
We emphasize that this is the first constraint coming from the analysis
of a single CMB measurement. Previous constraints on
$\alpha$ used a combination of various CMB and other cosmological observables, such as
supernovae, nucleosynthesis or the Hubble constant, in order to impose
stronger priors on other key cosmological parameters.
In this sense, the current WMAP data alone is at least as good as everything
else before it put together. As in previous work \cite{Old,Avelino,Martins}, 
the current data is
consistent with no variation, though the likelihood is skewed
towards smaller values at the epoch of decoupling.

We have shown that CMB data alone will be able to constrain $\alpha$ 
up to the  $0.1 \%$ level. Tighter constraints than this will require
invoking further (non-CMB) priors.
Nevertheless, the existence of an early reionization epoch as suggested by WMAP
turns out to be a blessing for these measurements, once more accurate cosmic microwave
background polarization data is available, and in particular we have proposed
a novel way of exploiting it.
These points are discussed in more detail in \cite{Rocha}.

\section{acknowledgments}
C.M. is funded by FCT (Portugal), under grant FMRH/BPD/1600/2000.
G.R. is funded by the Leverhulme trust.
This work was done in the context of the European network CMBnet, and
was performed on COSMOS, the Origin3800 owned by the UK
Computational Cosmology Consortium, supported by Silicon Graphics/Cray
Research, HEFCE and PPARC.


\bibliography{alphawmap}

\end{document}